\documentstyle[11pt,paspconf,epsf]{article}
\begin{document}

\title{Infrared Emission from AGN}

\author{D. B. Sanders}
\affil{Institute for Astronomy, University of Hawaii, 2680 Woodlawn Drive, 
       Honolulu, HI  96822}

\begin{abstract}
Infrared observations of complete samples of active galactic nuclei (AGN) have
shown that a substantial fraction of their bolometric luminosity is emitted at
wavelengths $\sim$8--1000$\mu$m.  In radio-loud and Blazar-like objects much of
this emission appears to be direct non-thermal synchrotron radiation.  However,
in the much larger numbers of radio-quiet AGN it is now clear that thermal dust
emission is responsible for the bulk of radiation from the near-infrared
through submillimeter wavelengths.  Luminous infrared-selected AGN are often
surrounded by powerful nuclear starbursts, both of which appear to be fueled by
enormous supplies of molecular gas and dust funneled into the nuclear region
during the strong interaction/merger of gas rich disks.  All-sky surveys in the
infrared show that luminous infrared AGN are at least as numerous as
optically-selected AGN of comparable bolometric luminosity, suggesting that AGN
may spend a substantial fraction of their lifetime in a dust-enshrouded phase.
The space density of luminous infrared AGN at high redshift may be sufficient
to account for much of the X-Ray background, and for a substantial fraction of
the far-infrared background as well. These objects plausibly represent a major
epoch in the formation of spheroids and massive black holes (MBH).
\end{abstract}



\section{Introduction}
Infrared observations\footnote{``Infrared" is used here to include rest-frame
emission over the broad wavelength range $\sim$8--1000$\mu$m (i.e. the
mid-infrared, far-infrared and submillimeter, but not the near-infrared).
Definitions for observed quantities such as $L_{\rm ir}$, $L_{\rm B}$, etc. are
taken from Table 1 of the review by Sanders \& Mirabel (1996). $H_{\rm o} = 75$
and $q_{\rm o} = 0$ is assumed throughout this article.} of AGN historically,
have lagged behind observations at shorter and longer wavelengths, therefore it
is not surprising that much of the literature is still biased toward studies of
radio and optical/X-Ray selected AGN.  While radio-loud objects and the highly
variable optical/X-Ray sources (e.g. Blazars, OVVs) provide an opportunity for
studying the physics of AGN, they draw attention away from the much larger
fraction of radio-quiet and dust-enshrouded sources.  These dusty AGN appear to
hold important clues for understanding the origin and evolution of all AGN, and
the relation of AGN to other classes of extragalactic objects.

This review focuses on the relatively large body of infrared continuum
data\footnote{Mid- and far-infrared spectroscopy of AGN is not covered in this
review.  Until very recently, such data were available for only a small number
of optically selected targets.  However, data for a larger number of nearby AGN
are now available from the {\it Infrared Space Observatory} ({\it ISO}), and
the reader is referred to articles by O. Laurent and J. Clavel at this
conference, and to the excellent paper by Genzel et al. (1998).} that is now
available for complete samples of nearby AGN, and stresses new results that
have been published since the previous IAU Symposium on AGN in Geneva, 1993
(``Multiwavelength Continuum Observations of AGN", S159).  A major highlight of
the most recent work is the clear identification of a large population of
dust-enshrouded AGN that may be more numerous than optically selected AGN in
the Universe.

In reviewing the infrared continuum properties of AGN, it is instructive to
trace the highlights of infrared studies from the first mid- and far-infrared
measurements of selected nearby targets in the late 60's and 70's, through to
the latest spacecraft results that bear directly on the AGN population in the
more distant Universe.  Infrared observations took a great step forward
following the all-sky surveys carried out by the Infrared Astronomical
Satellite ({\it IRAS}), and it seems natural to divide our initial discussion
accordingly, and then to show how infrared studies of AGN in the local Universe
can help in understanding the AGN population at high redshift.

\section{ Pre-{\it IRAS}: A Brief Historical Review, 1968-83}

The pioneering infrared observations of Low \& Kleinmann (1968), and Kleinmann
\& Low (1970a,b), followed by more accurate photometry by Rieke \& Low (1972),
made it clear that strong infrared emission could dominate the spectral energy
distributions (SEDs) of Seyfert galaxies, and even singled out a class of
objects with ``ultra-high" infrared luminosities that rivaled the bolometric
luminosity of QSOs.  The first clear evidence that the mid-infrared emission
from most Seyferts might not be direct synchrotron radiation was provided by
new 10$\mu$m data for NGC1068 which showed lack of variability (Stein et al.
1974), plus an extended source (Becklin et al. 1973).  The infrared spectrum
appeared to be better explained by models of thermal reradiation from dust
(e.g. Rees et al. 1969; Burbidge \& Stein 1970).

Mid-infrared ground-based photometry of large samples of Markarian Seyferts and
starbursts (Rieke \& Low 1975; Neugebauer et al. 1976), Seyfert galaxies (Rieke
1978), plus mid- and far-infrared observations of a few nearby Seyferts with
the Kuiper Airborne Observatory (Harper \& Low 1973; Telesco \& Harper 1980)
proved that ``infrared-excess" was a common property of Seyferts as well as
starburst galaxies, and with the possible exception of radio-loud objects and
QSOs, that this emission could indeed be understood in terms of thermal
emission from dust.  Rieke (1978) perhaps summarized it best (from a study of
50+ Markarian Seyferts) by stating that ``strong infrared excess is a virtually
universal characteristic of these sources.  ...the infrared continuum of a
number of type 1s and most type 2s is dominated by thermal reradiation by
dust", while also pointing out that the strength of the infrared excess was
correlated with the strength of the reddening (as measured by
H$\alpha$/H$\beta$), and that the proportionality between the 10$\mu$m and 21cm
fluxes noted earlier (e.g. Rieke \& Low 1972) was confirmed.

\begin{figure}
\plotfiddle{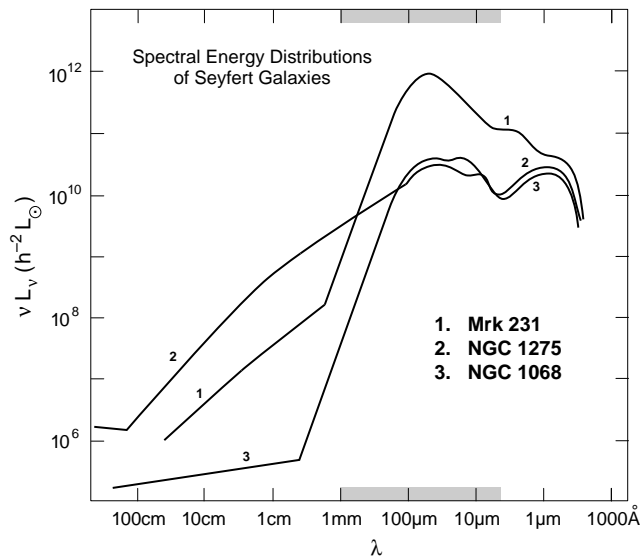}{2.5in}{0}{50}{50}{-160}{-80}
\caption{Spectral energy distributions from UV to radio wavelengths for 
three well-known AGN: the Sy1 galaxy Mrk231, the Sy2 galaxy NGC1068, 
and the Sy2 powerful radio galaxy NGC1275 (Perseus~A).}
\label{fig-1}
\end{figure}

Figure 1 shows SEDs using the most recent radio-to-UV continuum data for three
``classic" Seyferts, which were among the first AGN observed in the mid- and
far-infrared.  The steep submillimeter spectral index, $\alpha >$2.5 (where
$f_\nu \propto \nu^\alpha$), for radio-quiet objects confirms earlier
suggestions that the ``infrared bump" at wavelengths $\sim$5--500$\mu$m is due
to thermal emission from dust.  Only in the radio-loud sources such as NGC1275
(Perseus~A) is there any reason to believe that a substantial portion of the
infrared emission is simply the short wavelength extension of the non-thermal
emission seen at radio-to-millimeter wavelengths (e.g. Edelson \& Malkan
1986).

\section{Post-{\it IRAS}: Infrared Properties of Optical Samples of AGN}

{\it IRAS} was the first telescope with sufficient sensitivity to detect large
numbers of extragalactic sources at mid- and far-infrared wavelengths
(Neugebauer et al. 1984).  {\it IRAS} surveys of optically selected Seyfert
galaxies (Miley et al. 1985) and QSOs (Neugebauer et al. 1985, 1986) confirmed
that active galaxies could be strong infrared emitters; most optically selected
AGNs had ratios $L_{\rm ir}/L_{\rm B}$ in the range 0.2 to 1.0 with higher
values in only a small number of objects.

\subsection{QSOs}

\begin{figure}
\plotfiddle{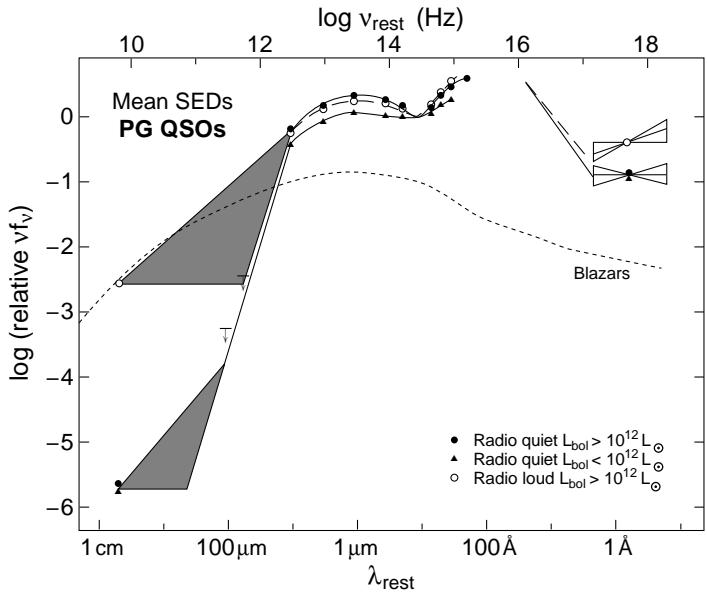}{3in}{0}{90}{90}{-280}{-250}
\caption{Mean spectral energy distributions from X-ray to radio wavelengths 
for optically selected radio-loud and radio-quiet QSOs (Sanders et al. 1989) 
from the Palomar-Green Bright QSO Survey (Schmidt \& Green 1983), and for 
Blazars (Impey \& Neugebauer 1988).}
\label{fig-2}
\end{figure}

A more complete accounting of the infrared properties of optically selected
Sy1s and QSOs is given by the data in Figure 2. The interpretation by Sanders
et al. (1989) was that the gross shape of the SEDs between 3000\AA\ and
300$\mu$m is remarkably similar for all QSOs (except the flat-spectrum
radio-loud quasars like 3C273) and that this can broadly be interpreted by two
broad components of thermal emission; the ``big blue bump" representing
$10^5$--$10^6$~K thermal emission from an accretion disk, and an ``infrared
bump" made up of reradiation from dust in a distorted disk extending from
$\sim$0.1~pc to more than 1~kpc.

Only for flat-spectrum radio-loud QSOs is there good evidence that much of the
infrared emission is probably direct non-thermal emission from the central
AGN.  These objects also tend to exhibit variability on relatively short
time-scales (hours to weeks).  In the highly variable Blazars there is strong
evidence that infrared emission is truly just part of a single non-thermal
spectral component from millimeter to optical wavelengths (see Figure 2).  The
range of variability in both the spectral index and flux density for Blazars
increases with decreasing wavelength, with the variability in the far-infrared
being less than half that observed in the optical.  However, a substantial
fraction of Blazars, when in their minimum variable state, also show evidence
for underlying emission lines and thermal infrared components, suggesting that
these objects may still contain substantial amounts of dust.

\subsection{Narrow-line AGN}

A more complete accounting of the infrared properties of optically selected
narrow-line AGN (NLAGN) is given by the data in Figure 3, which suggest a
larger mean infrared excess ($\equiv L_{\rm ir}/L_{\rm B}$) than previously
assumed; the mean value is $\sim$1 for LINERs and $\sim$5 for Sy2s.  Both
Seyferts and LINERs have a far-infrared peak at $\sim$60--200$\mu$m
characteristic of relatively ``cool" dust ($T_{\rm dust} \sim$25--50K), with
the strength of the infrared emission in Sy2s actually being somewhat larger
than that observed in starbursts (!), primarily due to added emission from a
second ``warm" ($T_{\rm dust} \sim$100--500K) dust peak in Sy2s at wavelengths
$\sim$5--50$\mu$m.

\begin{figure}
\plotfiddle{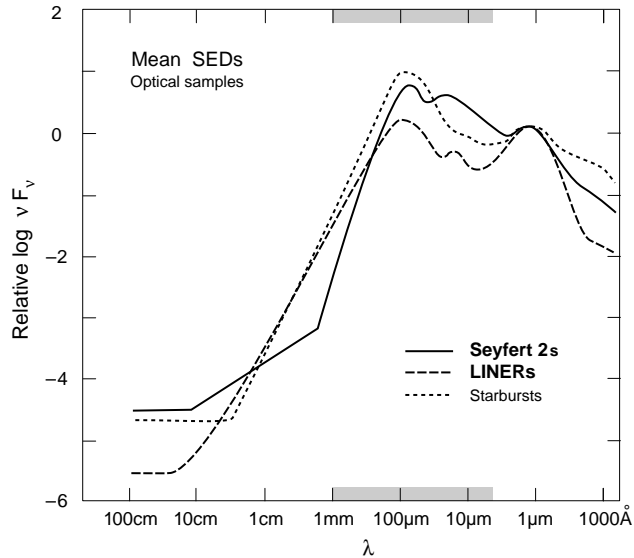}{2.5in}{0}{50}{50}{-160}{-100}
\caption{Mean spectral energy distributions from X-ray to radio wavelengths 
(normalized to log~$\nu F_\nu (7700\AA) \equiv 0$) for optically selected 
NLAGN (Sy2s and LINERs) compared with luminous starbursts (adapted from 
Schmitt et al. 1997).}
\label{fig-3}
\end{figure}

\section{Infrared Selected AGN}

The true extent of infrared excess exhibited by all types of extragalactic
objects was only begun to be realized following extensive ground-based
follow-up studies of complete infrared selected samples.  deGrijp et al.
(1985) found that searches based on ``warm" infrared colors ($f_{25}/f_{60} \ga
0.3$) could be useful for discovering {\it new} infrared-luminous AGN, a
technique that appeared to be motivated by the shape of the infrared spectrum
of the Sy2 galaxy NGC1068 (Telesco \& Harper 1980), and the discovery of a
similar ``warm" component in the broad-line, infrared-luminous radio galaxy
3C290.3 (Miley et al. 1984).  These infrared-selected AGN had a mean ratio of
$L_{\rm ir}/L_{\rm B} \sim 10$ and included a few objects with ratios as large
as $\sim$30--50.  deGrijp et al. (1985) suggested that the true space density
of AGNs could be a factor of two larger than previously assumed (with the
majority of the new infrared-selected objects being a mixture of Sy2s and
LINERs).

The first infrared-selected ``bonifide" QSOs \footnote{(i.e. Sy1s with $L_{\rm
ir} > 10^{12} L_\odot$; equivalent to the bolometric luminosity of optically
selected QSOs with $M_{\rm B} < -23$)} were also found during color-selected
searches of the {\it IRAS} database at the faintest flux levels (Beichman et
al. 1986; Vader \& Simon 1987; Low et al. 1988; Sanders et al. 1988b).
Additionally, the most luminous infrared sources (with $L_{\rm ir} \sim 10^{13}
L_\odot$), all of which were found to be dusty Sy2s in direct emission (e.g.
Kleinmann \& Keel 1987; Hill et al. 1987; Frogel et al. 1989; Cutri et al.
1994) were subsequently shown to be obscured infrared QSOs (i.e. Sy1s) in
polarized light (e.g.  Hines et al. 1995).  Whether all infrared selected Sy2s
harbor obscured Sy1s is not yet clear.  However, it seems plausible that
``unified models" invoked to understand the polarization properties of
optically selected AGN (e.g.  Antonucci \& Miller 1985), which suggest that the
observed spectral type (Sy1 vs. Sy2) depends largely on the orientation of a
circumnuclear dust torus to the line of sight (see Antonucci 1993 for a more
complete review) could play a major role in infrared-selected AGN.  Models with
even larger dust shrouds ($>$100 pc) where the obscuring material covers most
of the sky as seen from the central source (e.g. Fabian et al. 1998) may also
need to be invoked to account for the objects with the largest infrared
excess.

\subsection{AGN versus Infrared Luminosity}

\begin{figure}
\plotfiddle{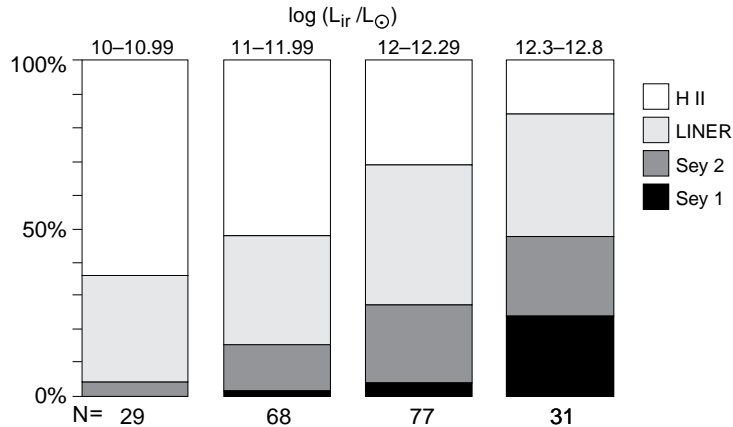}{2in}{0}{120}{120}{-370}{-425}
\caption{The optical spectral classification of flux-limited (60$\mu$m) 
samples of infrared selected galaxies versus infrared luminosity 
(Veilleux et al. 1999).}
\label{fig-4}
\end{figure}

Perhaps the most important spectroscopic result from studies of flux-limited
samples of extragalactic infrared objects it the increasing fraction of AGN
among the most luminous sources.  It is clear from Figure 4 that the fraction
of Seyferts increases systematically with increasing $L_{\rm ir}$, to where
{\it Seyferts account for nearly half of all objects at the highest infrared
luminosities}.  The ratio of Sy1s to Sy2s also increases to the point where
both are $\sim$25\% of the total number of objects with $L_{\rm ir} > 10^{12.3}
L_\odot$.  Whereas from optical surveys alone it had been thought that Sy2s
were very rare at high bolometric luminosities, it now seems clear that high
luminosity Sy2s were simply hiding as a subset of the most luminous infrared
selected galaxies.  Likewise for luminous H~II galaxies, although the fraction
of H~II galaxies diminishes to $\la$25\% for ultraluminous infrared galaxies
(ULIGs) at $L_{\rm ir} > 10^{12} L_\odot$.  LINERs appear to remain constant at
$\sim$1/3 of the sample at all $L_{\rm ir}$, and although LINERs have sometimes
been lumped together with NLAGN, recent evidence suggests that the emission
lines in many of these infrared-selected LINERs may be powered primarily by
shocks and superwinds from massive stars (e.g. Veilleux et al. 1999).

\section{The Starburst-AGN Connection}

\begin{figure}
\plotone{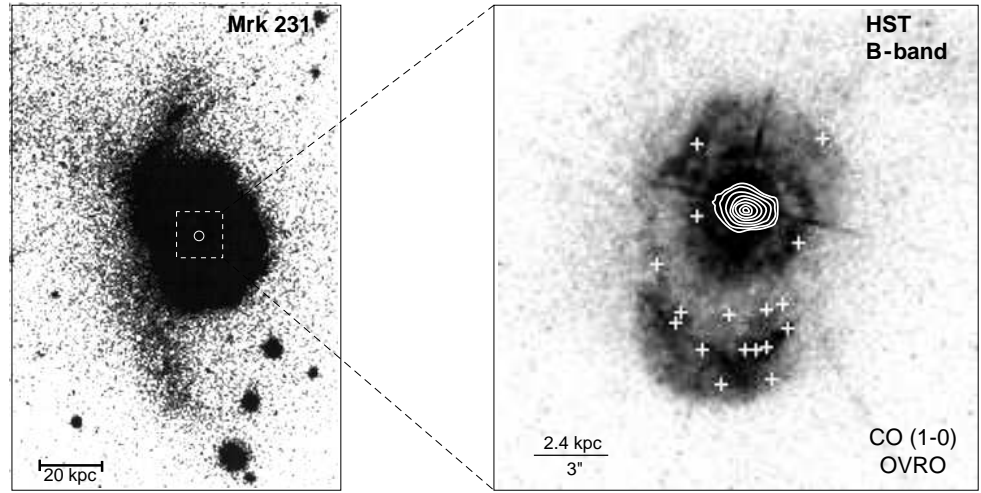}
\caption{The advanced merger/ULIG/QSO Mrk 231 -- Left panel: optical image 
(Sanders et al. 1987) and CO contour (Scoville et al. 1989).  Right panel: 
HST B-band image and identified stellar clusters (`+') from Surace et al. 
(1998).  The high resolution CO contours are from Bryant \& Scoville (1997).}
\label{fig-5}
\end{figure}

There is increasing direct evidence that powerful circumnuclear starbursts and
AGN may be intimately related. An excellent review of this subject can be found
in the Taipei Workshop - Relationships between AGN and Starburst Galaxies
(1992).  For the current discussion, the case of Mrk231 (Figure 5) is
instructive.  Mrk231, like other ULIGs, contains a large population of
relatively unobscured luminous star clusters at galactocentric radii
$\sim$0.5--3kpc (Surace et al. 1998).  The inner 1kpc region still contains an
enormous supply of gas and dust, much of which may be arranged in a
subkiloparsec disk (Carilli et al. 1998).  A partial face-on orientation for
this disk may be the explanation for why we can see the redenned Sy1 nucleus.

It has been suggested that both the intense circumnuclear starburst and the AGN
currently contribute approximately equally to the bolometric infrared
luminosity in Mrk231 (D. Weedman and H.E. Smith, this conference).  However,
the nature of the dominant power source for ULIGs continues to be the subject
of great debate (e.g. Ultraluminous Galaxies: Monsters or Babies, Ringberg
Workshop, 1998) with arguments favoring both circumnuclear starbursts and AGN
as well as nearly equal mixtures of the two phenomena.  Extensive
multiwavelength spectroscopic studies are currently underway in an attempt to
resolve the issue (e.g. Genzel et al. 1998), but due to heavy dust obscuration
it is not yet clear which process dominates in all objects.

\subsection{The Origin and Evolution of Infrared-luminous AGN}

Extensive ground-based observations of complete samples of infrared-selected
galaxies now clearly show that strong interactions/mergers of gas-rich spirals
play a dominant role in triggering the most luminous infrared systems (see the
review by Sanders \& Mirabel 1996).  The fraction increases from $\sim$30\% at
$L_{\rm ir} = 10^{11} L_\odot$ to $\ga$95\% for ULIGs at $L_{\rm ir} >10^{12}
L_\odot$ (e.g. Sanders et al. 1988a; Mirabel et al. 1990; Kim 1995; Murphy et
al. 1996; Clements et al. 1996).  Prominent tidal tails, similar to what is
seen for Mrk231 (Figure 5), are indeed found in all nearby ULIGs, suggesting
that the mergers involve two relatively large gas-rich spirals.  Figure 6
illustrates the ubiquitous tidal tails, and in some cases double nuclei, that
can still be detected in deep images of even more distant luminous infrared
objects, in this case three optically selected AGN.

\begin{figure}
\plotone{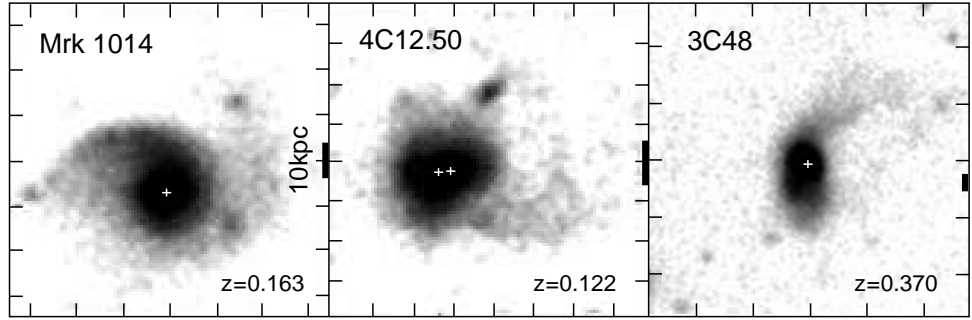}
\caption{Optical images of infrared-excess, optically selected QSOs, powerful 
radio galaxies, and infrared selected QSOs (MacKenty \& Stockton 1984; 
Kim 1995; Stockton \& Ridgway 1991).  The `+' sign indicates the position of 
putative optical nuclei. Tick marks are at 5$^{\prime\prime}$ intervals and 
the scale bar represents 10~kpc. All three objects exhibit strong nuclear 
concentrations of molecular gas, with typically $\sim 10^{10} M_\odot$ 
concentrated at galactocentric radii $\la$1~kpc (Sanders et al. 1988c; 
Mirabel et al. 1989; Scoville et al. 1989).}
\label{fig-6}
\end{figure}

Strong interactions/mergers of gas-rich spirals appear to be  extremely
efficient at funneling large amounts of gas into the merger nuclei (e.g. Barnes
\& Hernquist 1992; Mihos \& Hernquist 1994).  Kormendy \& Sanders (1992) have
summarized the evidence that these objects are elliptical galaxies in
formation. Intense starbursts are clearly involved in producing the bulk of the
infrared luminosity throughout much of the initial stages of this process, and
may continue to do so through the most intense infrared phase, although it
seems reasonable to assume that during the peak infrared phase (which is close
in time to when the two nuclei merge and also corresponds to the most compact
concentration of gas in the nuclear region), conditions would be most optimum
for building and fueling an AGN, and that the AGN may rival if not dominate the
luminosity output of the system.  Mrk231 and most other ULIGs would seem to be
at this stage.

Eventually, powerful superwinds (e.g. Armus et al. 1989), that are indeed
observed in ULIGs, may clear away much of the surrounding nuclear dust shroud.
It seems reasonable that this housecleaning process could terminate much of the
circumnuclear starburst (as it already may have in the 1-3 kpc annulus of
Mrk231), but that the fueling of the AGN may continue due to the strong
self-gravitation of the inner accretion disk surrounding the AGN.  This later
period may still be marked by infrared excess determined by conditions on much
smaller scales such as the thickness, orientation, and opening angle of the
dust and gas torus thought to surround the AGN.  Eventually the ``big blue
bump" normally associated with optically selected QSOs should emerge, as has
happened already in the case of Mrk1014 and 3C48 (Figure 6).  This scenario
would seem to provide a plausible explanation for a correlation between black
hole mass and bulge mass as recently found in nearby ellipticals (Kormendy \&
Richstone 1995) since the mass of both the black hole and spheroid may show
similar dependence on the total mass of gas funneled into the merger nucleus.

\section{Deep Infrared and X-Ray Surveys: AGN and the Distant Universe}

There is now good evidence that both ULIGs and NLAGN may be sufficiently
numerous at high redshift ($z >$1) to account for the infrared background and
the X-Ray background respectively.  There is also increasing circumstantial
evidence that suggests NLAGN may be a substantial subset of these high-redshift
ULIGs, which if true, would suggest that a major epoch in the formation of
spheroids and MBH may have finally been discovered.

The evidence from the X-Ray side is as follows.  Deep X-Ray surveys with ROSAT
are consistent with a space density that evolves as steeply as $(1+z)^5$ for
high-luminosity AGN out to at least $z \sim$2 with a constant value at higher
redshift (e.g. Hassinger et al. 1998).  These sources are sufficiently numerous
to account for nearly all of the observed X-Ray background; however, the shape
of the X-Ray background spectrum implies that most of these AGN are heavily
absorbed (e.g. Fabian \& Barcons 1992; Boyle et al. 1995; Almaini et al.
1998).  These ``narrow-line X-Ray galaxies" (NLXGs: Hassinger 1996) have
recently been characterized by Maiolino et al. (1998) as having ``extremely
heavy obscuration along the line of sight, ($N_{\rm H} > 10^{25}$cm$^{-2}$) in
most cases".

The evidence from the infrared side is as follows.  Evidence for space density
evolution as steep as $(1+z)^5$ is consistent with a series of studies at
varying far-infrared flux levels and wavelength bands.  In the mid- and
far-infrared, the deepest surveys carried out by {\it IRAS} (e.g. Hacking \&
Houck 1987; Lonsdale \& Hacking 1989; Gregorich et al. 1995; Kim \& Sanders
1998), and more recently the surveys with {\it ISO} (e.g. Taniguchi et al.
1997; Kawara et al. 1998; Aussel et al. 1998; Puget et al. 1998) are consistent
with evolution at least as steep as $(1+z)^5$ out to $z \sim$1.  Within the
past year, submillimeter surveys with the Submillimeter Common User Bolometer
(SCUBA) on the James Clerk Maxwell Telescope (Smail et al. 1997; Hughes et al.
1998; Barger et al. 1998; Eales et al. 1998) have revealed a substantial
population of ULIGs, consistent with steep evolution out to at least $z
\sim$2--3, and with constant space density at higher redshift.  These high
redshift ULIGs, which are almost certainly the high-$z$ extension of the
sources detected by {\it IRAS} and {\it ISO}, are sufficiently numerous to
account for {\it all} of the far-infrared/submillimeter background (e.g. Barger
et al. 1999).

To test the relationship between ULIGs and NLXGs at high redshift requires
sufficiently sensitive X-Ray and far-infrared/submillimeter surveys in
overlapping regions of the sky.  These data should be available within the next
few years from new X-Ray satellites and submillimeter interferometers.  For now
we can only note the interesting result that all of the ULIGs uncovered by {\it
IRAS} at $z >0.4$ (all of which have $L_{\rm ir} \sim 10^{13} L_\odot$) show
direct evidence for powerful AGN (e.g. Kleinmann \& Keel 1987; Rowan-Robinson
et al. 1991; Cutri et al.  1994), as does the first identified SCUBA source
(Ivison et al. 1998).  It would thus appear that the relationship shown in
Figure 4 indeed continues to higher luminosities and to higher redshift.

\section{Conclusions}

Infrared observations have shown clearly that thermal emission plays an
important, and often dominant, role in the total luminosity output of most
AGN.  Although in some objects -- most notably the flat-spectrum, radio-loud
AGN and the highly variable Blazars
 -- a substantial fraction of the observed infrared emission appears to be
direct non-thermal synchrotron radiation, the much larger number of radio-quiet
Sy1s and NLAGN appear to have infrared SEDs dominated by thermal emission from
dust.

The discovery by {\it IRAS} of a substantial population of infrared-selected
AGN suggests that a substantial fraction of the energy produced by accretion
may be absorbed by dust.  Ground-based observations of these dusty objects show
that powerful circumnuclear starbursts ($r \la$1kpc) are often closely linked
with the building and fueling of AGN, and that the most luminous sources
(ULIGs) are often associated with strongly interacting/merger galaxies.

There is now strong evidence that the space density of ULIGs was much larger in
the past ($z >$1), and that they may account for a large fraction, if not all,
of the far-infrared/submillimeter background.  If the trend of increasing AGN
fraction versus increasing infrared luminosity observed for local ULIGs
continues to high redshift, then dust-enshrouded AGN may account for much of
the X-Ray background as well.  This large population of high-redshift ULIGs may
represent an important stage in the formation of both spheroids and massive
black holes.

\acknowledgments

I am grateful to Karen Teramura for assistance in preparing the figures, and to
JPL contract no. 961566 for partial financial support.

\begin{question}{P. Veron}
You said that $\sim$30\% of IRAS galaxies are LINERs.  Many objects in 
the past have been called LINERs on the basis of insufficient spectroscopic 
data and probably have a composite spectrum (Sey2+H~II).  What 
definition of LINER have you used, and are your spectra always of good 
enough quality to classify without ambiguity those objects as LINERs ?
\end{question}
\begin{answer}{D. Sanders}
Approximately 30\% of the objects in Figure 4 have sufficient spectral 
coverage in the blue to include the O~II line, and for these, the LINER 
classification criteria are identical to the original definition given 
by Heckman (1980).  The remaining objects only have spectra covering 
$\sim$4000--8500\AA, and for these we have used the narrow-line AGN 
diagnostic diagrams of Veilleux \& Osterbrock (1987) with the less 
restrictive requirement that O~III/H$\beta$~$<$~0.5~.  All spectra are 
of sufficient quality to distinguish LINERs from either H~II or Sy2 
galaxies.  It is interesting to note that our finding that $\sim$1/3 
of infrared selected galaxies are LINERs (irrespective of infrared 
luminosity) is the same fraction found by Heckman (1980) for the 
percentage of LINERs among the ``nuclei of bright optically selected 
galaxies". 
\end{answer}

\begin{question}{P. Osmer}
What fraction of AGNs are being missed in optical samples (because of 
dust obscuration) ?
\end{question}
\begin{answer}{D. Sanders}
deGrijp et al. (1985) originally estimated that the total number of Sy2s 
would at least double based on the numbers of new Seyferts discovered in 
infrared color-selected samples. Our more recent results, obtained from 
ground-based spectroscopy of complete flux-limited samples selected at 
60$\mu$m (e.g. Veilleux et al. 1999), suggests a 2--4 fold increase in 
the number of luminous ($L_{\rm bol} > 10^{11.5} L_\odot$) Seyferts; 
most of these are Sy2s. 
\end{answer}

\end{document}